\documentclass[aps,prb,twocolumn,longbibliography,superscriptaddress,10pt]{revtex4-2}
\pdfoutput=1

\usepackage{amsmath}
\usepackage{amssymb}
\usepackage{subfigure}
\usepackage{dcolumn}
\usepackage{bm}
\usepackage{bbm}
\usepackage{amsthm}
\usepackage{thmtools}

\usepackage{mathtools}
\usepackage{physics}
\usepackage{binarytree}
\usepackage{verbatim}
\usepackage{xcolor}
\usepackage{tabularray}
\usepackage{tabularx}
\usepackage{times}
\usepackage[colorlinks,bookmarks=true,citecolor=green,linkcolor=blue,urlcolor=blue]{hyperref}
\usepackage{graphicx}

\begin{document}

% \title{Investigation of $J_1$-$J_2$ Heisenberg model on triangular lattice: A study with symmetric projected entangled-pair states}

\title{Investigation of the $J_1$-$J_2$ Heisenberg model on the triangular lattice: \\ A study with projected entangled-pair states}

\author{Litao Ma}
\affiliation{Center for Neutron Science and Technology, Guangdong Provincial Key Laboratory of Magnetoelectric Physics and Devices, School of Physics, Sun Yat-sen University, Guangzhou 510275, China}

\author{Wei-Lin Tu}
\affiliation{Graduate School of Science and Technology, Keio University, Yokohama, Kanagawa 223-8522, Japan}
\affiliation{Keio University Sustainable Quantum Artificial Intelligence Center (KSQAIC), Keio University, Tokyo 108-8345, Japan}

\author{Didier Poilblanc}
\affiliation{Laboratoire de Physique Th\'eorique, IRSAMC, Universit\'e de Toulouse, CNRS, UPS, France}

\author{Ji-Yao Chen}
\email{chenjiy3@mail.sysu.edu.cn}
\affiliation{Center for Neutron Science and Technology, Guangdong Provincial Key Laboratory of Magnetoelectric Physics and Devices, School of Physics, Sun Yat-sen University, Guangzhou 510275, China}

\date{\today}

\begin{abstract}
The nature of the quantum spin liquid (QSL) phase in the frustrated $J_1$-$J_2$ Heisenberg model on the triangular lattice remains an open and actively debated problem. In this work, we employ the infinite projected entangled-pair state (PEPS) to systematically investigate the model under different symmetry constraints. Our simulations reveal a direct transition from the $120^\circ$ N\'eel state to a putative QSL at $J_2/J_1\approx 0.08$, signaled by the collapse of magnetic order. We further show that, through either an appropriate unitary rotation or spontaneous spin long-range order, the stripe antiferromagnetic phase can also be accurately captured within the infinite PEPS framework. A central focus of our study is the role played by the PEPS symmetry in approximating the QSL ground-state sandwiched between the two magnetic phases. We first found that a fully-symmetric topological $\mathbb{Z}_2$ Resonating Valence Bond state, which can be written as a simple PEPS with bond dimension $D=3$, exhibits a reasonably good variational energy. Motivated by this finding, we have further constructed generic $\mathbb{Z}_2$-symmetric PEPS of larger bond dimension (up to $D=7$). We found that, under wavefunction optimization, spinons condense and, simultaneously, topological vison excitations get confined, hence precluding $\mathbb{Z}_2$ topological order. This strongly indicates the gapless (or critical) nature of the QSL phase, which is most naturally consistent with a U(1) Dirac spin liquid scenario.
% similar to the U(1) Dirac spin liquid phase. 

\end{abstract}

\maketitle

\section{Introduction}

A quantum spin liquid (QSL) is an exotic state of matter that evades long‑range magnetic order even in the thermodynamic limit~\cite{Savary2016}. Since the concept was first introduced by Anderson~\cite{Anderson1973}, the study of QSLs has inspired extensive theoretical and experimental efforts~\cite{Savary2016,Zhou2017}. Given how common simple ordered states are, one natural strategy for realizing QSLs is to look for mechanisms that destabilize such order, and frustration is therefore a key ingredient. In two dimensions, nearest‑neighbor interactions on kagome or triangular lattices provide canonical examples of geometric frustration~\cite{Savary2016}. Among these, the triangular lattice plays a particularly central role: even the simplest antiferromagnetic nearest‑neighbor Heisenberg model on this lattice is frustrated, since spins on each triangle cannot simultaneously satisfy all pairwise antiferromagnetic bonds. This inherent incompatibility produces a highly degenerate manifold of classical states and strongly enhances quantum fluctuations in the quantum case. Building on this foundation, one can introduce further‑neighbor couplings to tune the degree and character of frustration. A natural extension is the triangular‑lattice $J_1$–$J_2$ model, where antiferromagnetic second‑neighbor interactions compete with the nearest‑neighbor exchange. This competition enriches the phase diagram, destabilizes conventional magnetic order, and has long been viewed as a promising route toward realizing QSL behavior on the triangular lattice~\cite{Zhu2015,Hu2015,Iqbal2016,Gong2019,Hu2019}.

In recent years, advances in experimental techniques have enabled the identification of a growing number of materials exhibiting behaviors consistent with QSL on triangular lattices. Among these, $\mathrm{YbMgGaO}_4$ has stood out as a prominent candidate~\cite{Shen2016,Paddison2017,Li2017}. In this compound, $\mathrm{Yb}^{3+}$ ions with an effective spin‑1/2 form a triangular lattice, giving rise to strong geometric frustration. Neutron scattering measurements performed at ultra‑low temperatures have revealed a broad, continuum‑like magnetic excitation spectrum, one of the hallmark signatures of fractionalized excitations in a QSL state~\cite{Paddison2017}. More recently, $\mathrm{NaYbSe}_2$ has emerged as another compelling triangular‑lattice system. Its $\mathrm{Yb}^{3+}$ ions form an almost ideal triangular network and can be effectively modeled by a $J_1$–$J_2$ Heisenberg Hamiltonian~\cite{Scheie2024}, placing it in close proximity to theoretically proposed QSL regimes. A combination of experimental probes, including neutron scattering, pressure‑dependent ac calorimetry, and ac magnetic susceptibility, has provided converging evidence that $\mathrm{NaYbSe}_2$ hosts a QSL ground state~\cite{Scheie2024}. Together, these materials exemplify how triangular‑lattice rare‑earth compounds have become central platforms for exploring and validating QSL physics, spurring further theoretical and numerical studies.

\begin{figure}[htbp]
\centering 
\includegraphics[width=0.95\columnwidth]{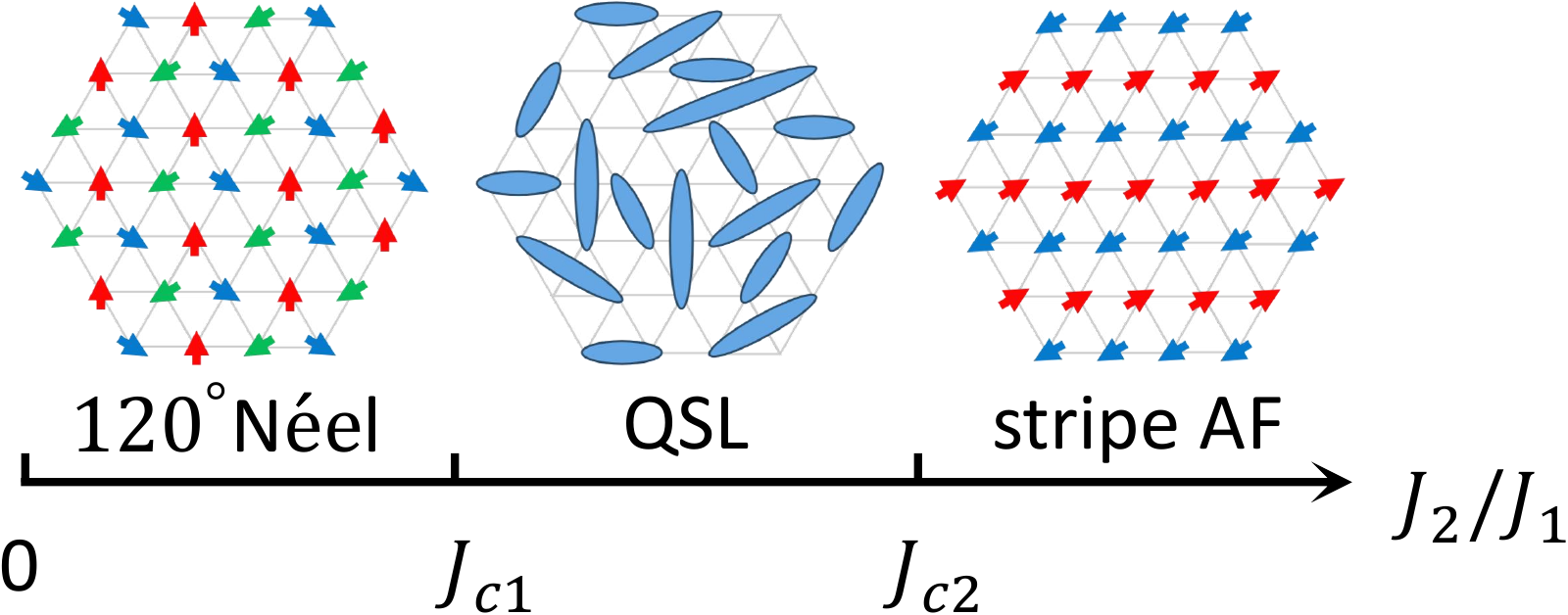} 
\caption{
Phase diagram of the $J_1$-$J_2$ model on triangular lattice.
For small $J_2/J_1$, the system stabilizes the $120^\circ$ N\'eel antiferromagnetic order. Upon increasing $J_2/J_1$, the first phase transition occurs at $J_{c1}$, where the magnetic order melts and the system enters a QSL phase. With further enhancement of $J_2/J_1$, the QSL regime eventually gives way to a stripe antiferromagnetic (stripe AF) phase, with the corresponding transition taking place at $J_{c2}$.
}
\label{Fig:phase_diagram}
\end{figure}

In fact, extensive numerical investigations have established the triangular‑lattice $J_1$–$J_2$ Heisenberg model as a fertile platform for realizing QSL behavior. Early variational Monte Carlo (VMC) studies revealed strong competition between magnetic order and spin‑liquid states in the intermediate‑coupling regime~\cite{Kaneko2014}. Large‑scale density matrix renormalization group (DMRG) calculations further demonstrated the emergence of a robust spin‑liquid phase, sandwiched by a $120^\circ$ Néel state with transition point at $J_{c1}$ and a stripe antiferromagnetic~(AF) state with transition point at $J_{c2}$, as indicated in Fig.~\ref{Fig:phase_diagram}~\cite{Zhu2015,Hu2015,Iqbal2016,Saadatmand2017}. It is important to emphasize that numerical studies have not yet reached a consensus on the precise nature of the quantum spin liquid in the triangular‑lattice $J_1$–$J_2$ model. VMC calculations generally suggest a gapless spin‑liquid state~\cite{Iqbal2016}, whereas DMRG simulations tend to support a gapped $\mathbb{Z}_2$ spin liquid~\cite{Zhu2015,Hu2015}. Recent high‑precision DMRG~\cite{Jiang2023} and dynamical DMRG~\cite{Jiang2026} study further reinforces the $\mathbb{Z}_2$ scenario. In contrast, spectral features obtained by matrix‑product‑state (MPS) calculations, large-scale exact diagonalization (ED), and VMC reveal features consistent with a gapless $\mathrm{U}(1)$ Dirac spin liquid~\cite{Hu2019,Ferrari2019,Sherman2023, Drescher2023,Wietek2024,Kovalska2026}. Note that experimentally, a very recent study of $\mathrm{NaYbSe}_2$ seems to indicate a low-energy gap, ruling out the gapless-spin-liquid scenario~\cite{Scheie2024}. Overall, these complementary approaches consistently identify a robust quantum‑disordered regime stabilized by the competition between nearest‑ and next‑nearest‑neighbor interactions. However, despite this agreement on the existence of a spin‑liquid phase, its detailed character, e.g. gapped $\mathbb{Z}_2$ versus gapless $\mathrm{U}(1)$, remains an open and actively debated question.

In this work, we employ the infinite projected entangled‑pair state (PEPS), a two‑dimensional tensor network ansatz~\cite{Verstraete2004,Cirac2021}, enhanced by the incorporation of various symmetry structures~\cite{Jiang2015,Mambrini2016,Chen2018_2,Chen2020}, to investigate the $J_1$–$J_2$ model on the triangular lattice, providing a new angle to approach this model. Our calculations reveal a clear phase transition in the ground state at a ratio $J_2/J_1\approx 0.08$, which aligns well with previous numerical studies. Although our infinite PEPS ansatz assumes translational invariance, we demonstrate that the stripe antiferromagnetic phase can also be captured either via (spontaneous) long-range order or using an appropriate unitary rotation to enlarge the unit cell, thereby showing that our approach is sufficiently flexible to capture multiple competing orders in this model.

A central aspect of our study is the incorporation of symmetries into the PEPS ansatz to examine how different symmetry constraints influence both the energetics and the nature of the optimized state in the QSL regime. We find that enforcing full $\mathrm{SU}(2)$ symmetry together with lattice point group symmetries, yields reasonable ground‑state energies. However, the high computation cost makes it prohibitive to increase the PEPS bond dimension. When restricting the tensors to only virtual $\mathbb{Z}_2$ symmetry, the variational energy exhibits a systematic decrease with bond dimension, reaching a value competitive to known results and indicating a more favorable variational description. Quite surprisingly, a careful examination of the properties of the optimized $\mathbb{Z}_2$ symmetric ansatz indicates spinon condensation and, hence, absence of $\mathbb{Z}_2$ topological order. Taken together, these results highlight both the versatility of our tensor‑network construction and the crucial role of symmetry implementation in resolving the nature of the QSL phase.

The plan of the paper is the following: first, we will present in Sec.~\ref{Sec:unrestricted_PEPS} results based on the unrestricted ansatz on a coarse-grained lattice, providing an accurate description of the two magnetic phases and the location of phase boundary between N\'eel phase and QSL phase in parameter space. In a second step, we shall highlight in Sec.~\ref{Sec:QSL_phase} the importance of choosing relevant symmetries in order to follow different paths to consistently describe the intermediate QSL phase. Sec.~\ref{Sec:conclusion} is devoted to the discussion and summary.

\section{Unrestricted PEPS on the coarse-grained lattice}
\label{Sec:unrestricted_PEPS}

\subsection{Model and method}

The spin-$1/2$ $J_1$-$J_2$ Heisenberg model on a triangular lattice is defined by the Hamiltonian:
\begin{equation}
H = J_1 \sum_{\langle ij \rangle} \mathbf{S}_i \cdot \mathbf{S}_j + J_2 \sum_{\langle\langle ij \rangle\rangle} \mathbf{S}_i \cdot \mathbf{S}_j,
\label{Eq:model}
\end{equation}
where $\langle ij \rangle$ and $\langle\langle ij \rangle\rangle$ denote nearest-neighbor (NN) and next-nearest-neighbor (NNN) bonds, respectively. Throughout this work we set $J_1=1$ being the energy unit.

\begin{figure}[htbp]
\centering 
\includegraphics[width=0.95\columnwidth]{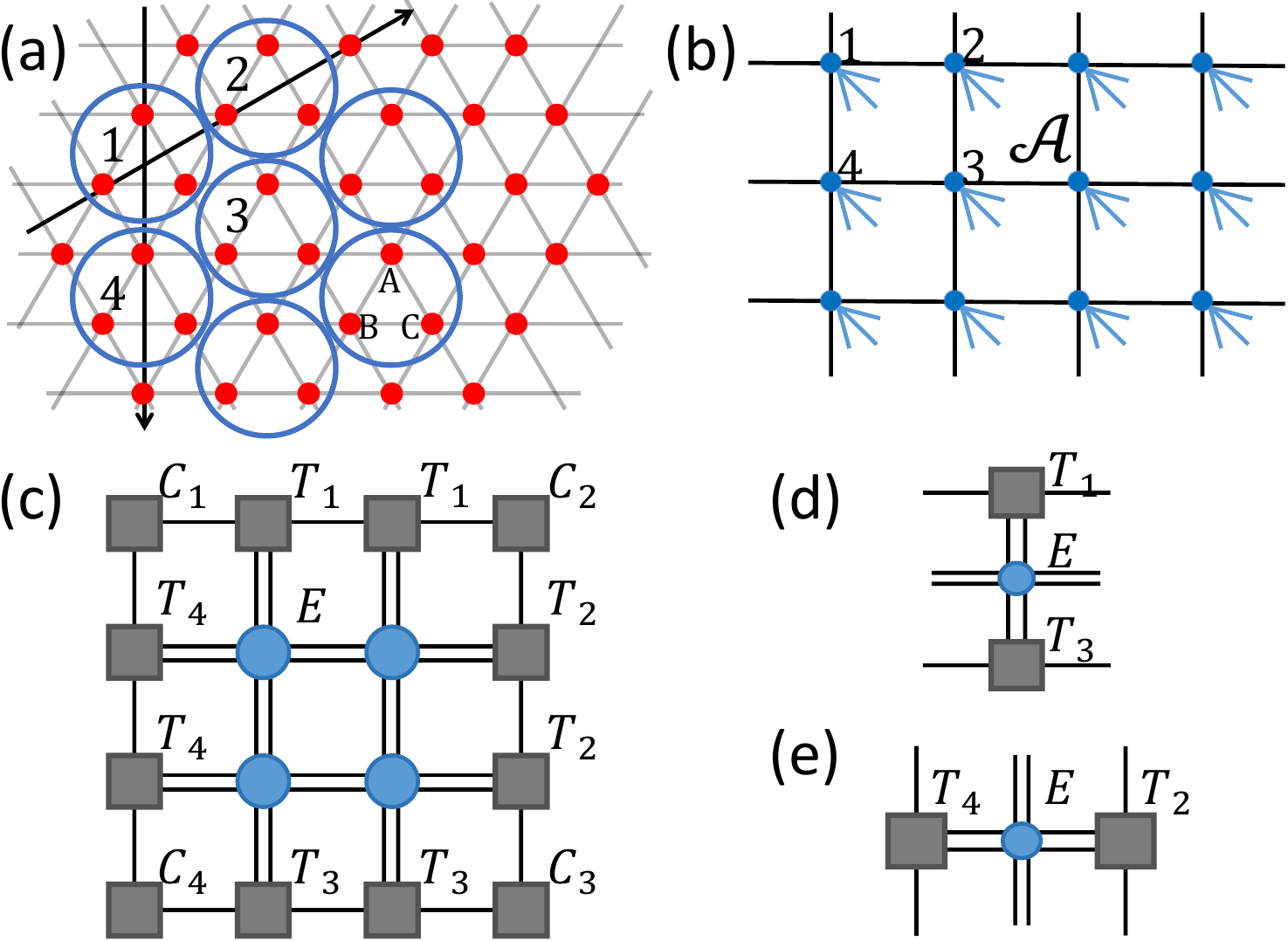}
\caption{PEPS setup on a coarse-grained lattice.
(a) Grouping three sites within each blue circle (denoted as $A,B,C$) on the triangular lattice, an effective square lattice is realized, with lattice translation vector indicated by arrows. (b) On the square lattice, we define the PEPS ansatz with single site unit cell tensor $\mathcal{A}$. The numbers in (a) and (b) show how the mapping is precisely performed, with $i$-th circle mapped to $i$-th site. (c) Denoting the double tensor as $E$, to compute observables with PEPS, the CTMRG method is used, producing environment tensors $\{C_i,T_i\},i=1,\ldots,4$ self-consistently. The same environment tensors are also used to compute correlation length along $x$,$y$ direction, using transfer matrix shown in (d) and (e), respectively. Note that the square lattice correlation length has to be multiplied by a factor $\sqrt{3}$ to obtain the value in terms of the triangular lattice spacing.
}
\label{Fig:PEPS_setup}
\end{figure}

Since each site on the triangular lattice has six nearest neighbors, a standard construction of PEPS on triangular lattice (see next section) will result in a rank-7 tensor, which is numerically expensive when considering relatively large bond dimension. Following Ref.~\cite{Chi2022}, to perform simulations with infinite PEPS, we map the triangular lattice onto a square‑lattice geometry, which is more convenient for numerical PEPS calculations. Figs.~\ref{Fig:PEPS_setup} (a), (b) illustrate the procedure, where we begin by marking the original lattice sites (indicated by red dots in Fig.~\ref{Fig:PEPS_setup}(a)) and then group three sites in each circle into one site. For latter convenience, we denote the three sites in each circle as belonging to sublattice $A,B,C$ of the triangular lattice, respectively. The resulting lattice can be viewed as an effective square lattice, on which we can define the PEPS with rank-5 tensors. To make the correspondence between degrees of freedom on the original triangular lattice and the coarse-grained lattice unambiguous, we have further shown the translation vector of the square lattice in Fig.~\ref{Fig:PEPS_setup}(a). As a result, the circles labeled as ``1,2'' and ``1,4'' become nearest neighbor sites (with the same label) in Fig.~\ref{Fig:PEPS_setup}(b), while ``1,3'' becomes next-nearest neighbor on the square lattice. Note that, although this reorganization is useful for numerical purpose, certain lattice point group symmetries are not explicit on the coarse-grained lattice, and thus a careful check would be needed to ensure a correct interpretation of the results.

For infinite PEPS simulations, it is essential to specify the size of the repeating unit cell. From the mapping procedure, it is clear that for the $120^\circ$ magnetic ordered phase and quantum spin liquid phase in the triangular $J_1$-$J_2$ model, a single‑site unit cell on the square lattice would be sufficient to represent the full infinite PEPS network. The stripe‑ordered phase constitutes an exception to the simple one‑site periodicity because of a $\pi$-phase shift between neighboring sites. This difficulty can be overcome by applying an appropriate rotation to the local tensor, which restores an effective one‑site unit‑cell description for the stripe phase as well~\cite{Vanderstraeten2016,Hasik2021}.

Denote the one-site unit cell tensor as $\mathcal{A}$, which is a rank‑5 tensor with four auxiliary bonds of dimension $D$ and one physical bond of dimension $d=2^3$, corresponding to the local Hilbert space of three spin‑1/2 degrees of freedom. The state is defined as
\begin{equation}
    |\Psi\rangle = \sum_{\mathbf{s}_1,\ldots,\mathbf{s}_N}\mathrm{Tr}\left(\mathcal{A}^{\mathbf{s}_1}\cdots \mathcal{A}^{\mathbf{s}_N}\right)|\mathbf{s}_1,\ldots,\mathbf{s}_N\rangle,
\label{Eq:CG_PEPS}
\end{equation}
where ``$\mathrm{Tr}$'' means identifying and summing over all virtual indices, and $\mathbf{s}_i$ represents the coarse-grained physical indices. The bond dimension $D$ serves as the key control parameter in PEPS calculations. From a computational perspective, increasing $D$ systematically improves the accuracy of the variational state but also leads to a rapid growth in numerical cost. Consequently, performing finite‑$D$ scaling and extrapolating observables obtained at several accessible bond dimensions toward the $D\rightarrow \infty$ limit are necessary for obtaining reliable thermodynamic‑limit estimates. In this section, we have carried out simulations on the coarse-grained triangular lattice with bond dimensions ranging from $D=2$ to $D=7$, using an unrestricted PEPS ansatz, where no further restriction on the local tensor $\mathcal{A}$ has been imposed except the tensor only contains real entries.

\begin{figure}[h]
\centering 
\includegraphics[width=0.97\columnwidth]{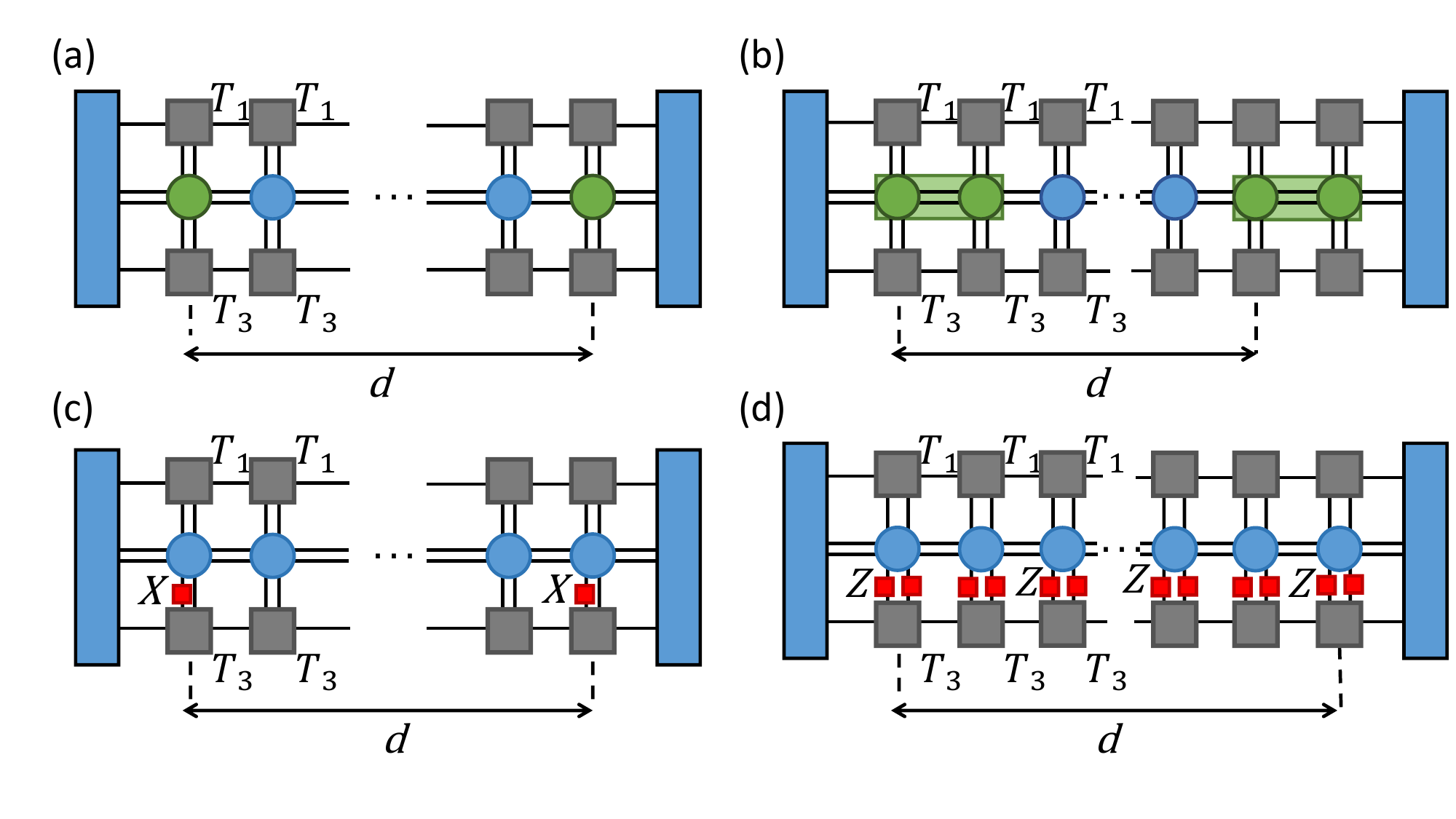}
\caption{
Setups for computing spin (a), dimer (b), spinon (c) correlation functions and wave function norm with inserted flux line (d). Operators acting on the physical (virtual) degrees of freedom are shown in green (red):
(a) Two local spin operators are acting on two sites at distance $d$. (b) Two pairs of NN spin operators are acting at distance $d$. (c) Two local $X$ operators are acting on two bonds (on a single layer) at distance $d$. 
(d) Two strings of $Z$ operators of length $d$ are acting on the bonds of both top and bottom layer. The last two setups (c) and (d) are used for $\mathbb{Z}_2$ symmetric PEPS (see Subsec.~\ref{Subsec:Z2}).
} 
\label{Fig:Z2corrlenfig}
\end{figure}

Another key component of the PEPS framework is the treatment of the environment tensors. To access the thermodynamic limit, we employ the corner transfer matrix renormalization group (CTMRG) procedure~\cite{Nishino1996,Orus2009,Corboz2014}. The first step is to contract each local tensor with its Hermitian adjoint along the physical index, producing a double‑layer tensor $E$ whose virtual bonds have dimension $D^2$. We then apply the CTMRG algorithm to iteratively build and renormalize the corner and edge tensors until convergence is reached. The resulting fixed‑point tensors serve as an effective ``bath'' that captures the influence of the infinite surrounding lattice on the bulk tensor. To efficiently incorporate next‑nearest‑neighbor interactions, we have used the setup shown in Fig.~\ref{Fig:PEPS_setup}(c) to compute the relevant expectation values. This setup ensures that all local observables, e.g., energy density, magnetization, and correlation length (see Figs.~\ref{Fig:PEPS_setup}(d)(e)) can be easily obtained from the iPEPS ansatz. The fixed-point edge tensors can also be used to compute various types of correlation functions as shown in Figs.~\ref{Fig:Z2corrlenfig}.

To obtain a faithful representation of the ground state of the target Hamiltonian, we variationally optimize the PEPS tensor $\mathcal{A}$ through a gradient based energy‑minimization scheme~\cite{Vanderstraeten2016,Corboz2016}, with gradient obtained by the backward mode of automatic differentiation~\cite{Liao2019}. After constructing the double‑layer tensor network and its corresponding CTMRG environment, the variational energy becomes an explicit differentiable function of all tensor elements. Automatic differentiation allows to compute energy gradients with respect to PEPS tensor $\mathcal{A}$, facilitating efficient updates through a conjugate gradient method~\cite{NoceWrig06}. By iteratively adjusting the tensor to lower the variational energy and repeatedly recomputing the environment until convergence, the infinite PEPS network is driven towards the optimal fixed point that approximates the ground‑state wave function in the thermodynamic limit.

\subsection{The Heisenberg limit}

We now present our PEPS results, where we have made sure that all data are converged with respect to the environment bond dimension in CTMRG. As an initial step, we establish a high‑precision benchmark for the ground state at $J_2=0$, a well‑studied limit that provides a stringent test of our numerical approach.

Within our unrestricted iPEPS framework, we compute both the ground‑state energy and coplanar onsite magnetization defined as
\begin{equation}
m^2=\frac{1}{3}\sum_{i=A,B,C}\left((m_i^x)^2 + (m_i^y)^2 + (m_i^z)^2\right ),
\label{eq:magnetization}
\end{equation}
where $m_i^\alpha=\langle S_i^\alpha\rangle$ denotes the $\alpha$-component of the local magnetic moment on sublattice $i=A,B,C$ of the triangular lattice (cf. Fig.~\ref{Fig:PEPS_setup}(a)). Although the ordering wave vectors differ, $\mathbf{k}=(2\pi /3,2\pi /3)$ for the $120^\circ$ N\'eel state and $\mathbf{k}=(\pi, 0)$ (and symmetry related vectors) for the stripe AF phase, ideally the magnitude of the ordered moment is uniform across all lattice sites for both magnetic ordered phases. Due to the mapping from the triangular lattice to the square lattice, the magnitude of the moment is not exactly the same on the three sublattices for the $120^\circ$ phase, though the differences are quite small. Thus we have taken an average in Eq.~\eqref{eq:magnetization}. Moreover, since we have used PEPS tensor with real elements, $m_i^y=0$ and the magnetizations on all three sublattices lie in the $xz$ plane. We have checked explicitly that the angles between ordering vectors are close to $120^\circ$ (with deviations smaller than $1^\circ$), even for the smallest bond dimension $D=2$.

\begin{figure}[htbp]
\centering 
\includegraphics[width=0.95\columnwidth]{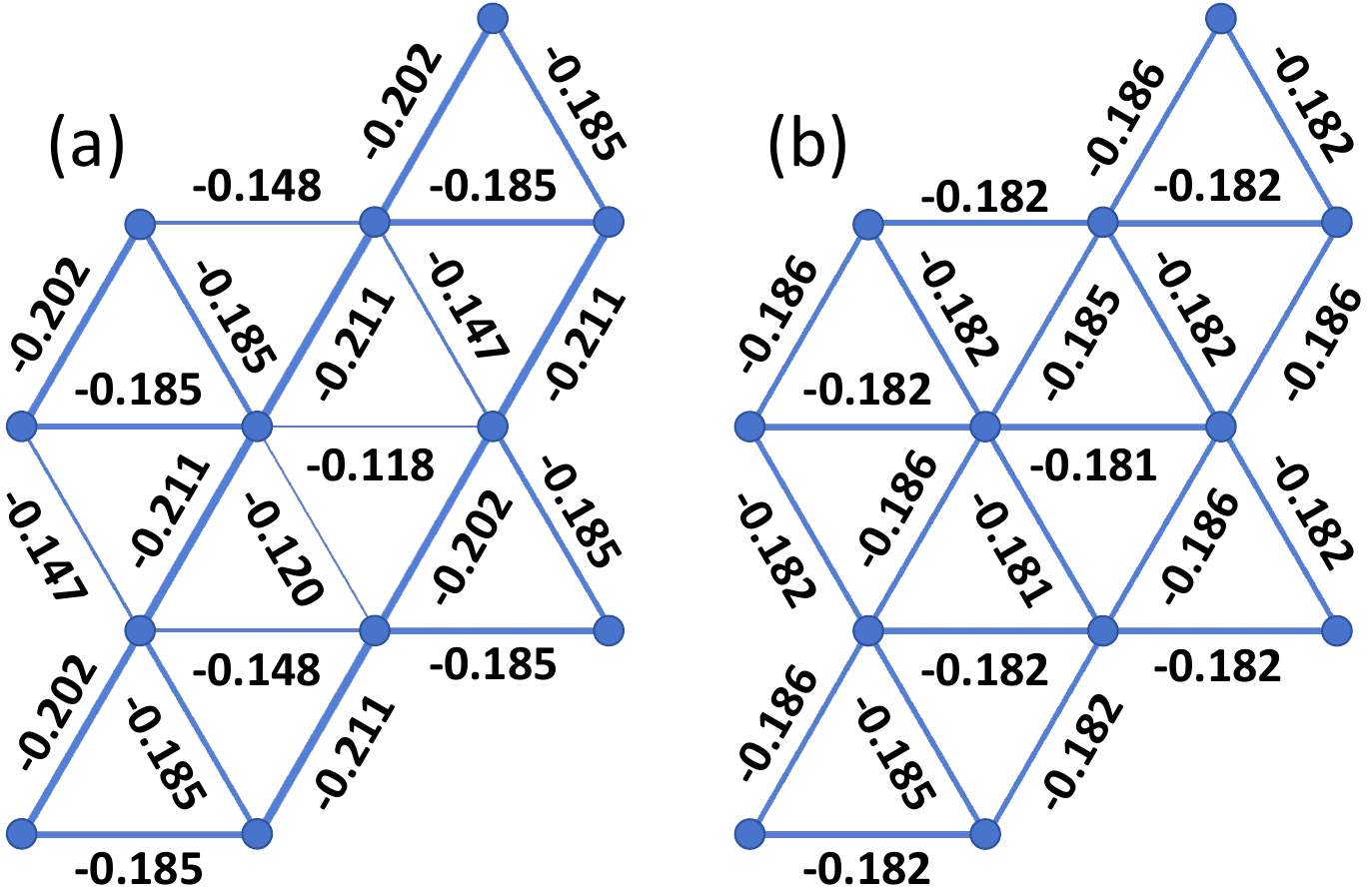}
\caption{Bond energies of the triangular Heisenberg model with $J_2=0$, computed using unrestricted iPEPS on the coarse-grained lattice. In both (a) and (b), the line widths indicate the absolute value of bond energies. Increasing bond dimension from (a) $D=2$ to (b) $D=6$, the lattice rotation symmetry is gradually restored.
}
\label{Fig:Edistribution}
\end{figure}

For the expectation values of Hamiltonian terms, the $C_6$ lattice rotation symmetry is broken at small bond dimension (e.g., $D=2$), but is gradually restored with increasing $D$. See Fig.~\ref{Fig:Edistribution} for an explicit comparison between $D=2$ and $D=6$. This provides confidence that this setup captures the correct physics, yielding reliable numerical results even at moderate $D$.

Since all results are computed at finite bond dimension, to get an estimate of physical quantities in the large bond dimension limit, we carry out a scaling analysis for the $J_2=0$ case. Two different scaling forms have been proposed and used in the literature: the finite bond dimension scaling and finite correlation length scaling~\cite{Corboz2018,Rader2018}. The latter is well suited for Lorentz invariant gapless systems. In our case, the Goldstone mode from spontaneous symmetry breaking of $\mathrm{SU}(2)$ symmetry renders this scaling form promising.

\begin{figure}[htbp]
\centering
\includegraphics[width=0.95\columnwidth]{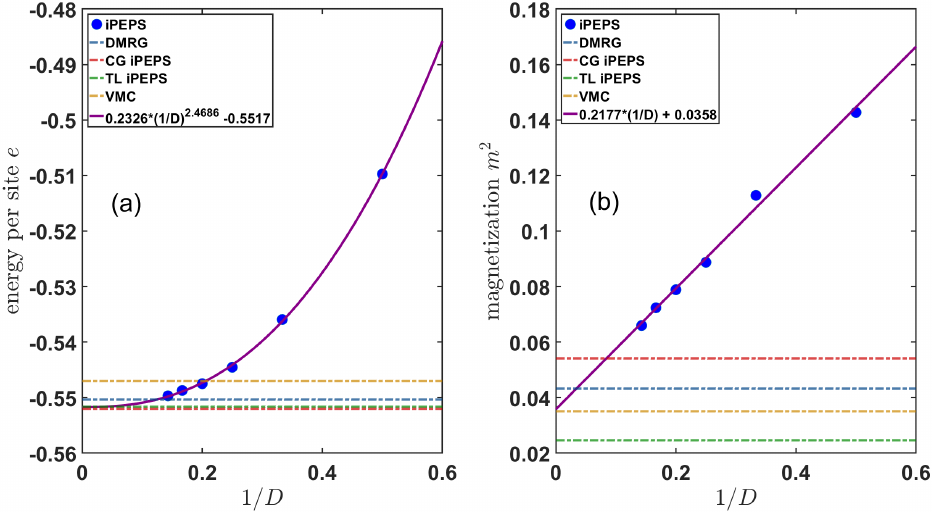}
\caption{Ground‑state energy per site $e$ and magnetic order parameter $m^2$ for the nearest‑neighbor Heisenberg model on the triangular lattice. The numerical data are extrapolated to the $1/D=0$ limit~(purple lines). Our extrapolated values are compared with results from DMRG~\cite{Huang2024}, coarse‑grained (CG) iPEPS~\cite{Chi2022}, triangular‑lattice (TL) iPEPS~\cite{Naumann2026}, and VMC~\cite{Heidarian2009}.
}
\label{Fig:finite_D_scaling}
\end{figure}

In Fig.~\ref{Fig:finite_D_scaling}, we carry out the finite bond dimension scaling analysis. We found that a power-law fitting in $1/D$ for the energy per site works well~\cite{Chi2022}, and similarly a linear $1/D$ fit for the magnetization $m^2$. In Ref.~\cite{Chi2022}, $m$ is used instead of $m^2$ in the linear fit versus $1/D$. Here we have checked that both $m$ and $m^2$ fittings work well, while for nonzero $J_2$, the latter produces more reasonable results. Extrapolating to the $D \to \infty$ limit provides the corresponding observables in the thermodynamic limit. This procedure yields the estimated energy density $e=-0.5517$ and magnetization $m=0.1892$. To assess the reliability of this approach, we compare these values with state‑of‑the‑art numerical results reported in the literature, as summarized in Table~\ref{Tab:benchmark}. Our extrapolated values show excellent agreement with DMRG~\cite{Huang2024}, coarse‑grained~(CG) iPEPS~\cite{Chi2022}, triangular‑lattice~(TL) iPEPS~\cite{Naumann2026} studies, and VMC~\cite{Heidarian2009}.

\begin{table}[htbp]
\centering
\caption{Comparison of the ground‑state energy per site $e$ and the magnetization $m^2$ obtained from different numerical methods for the triangular lattice Heisenberg model ($J_2=0$).}
\label{Tab:benchmark}
\begin{tabularx}{0.48\textwidth}{lll}
\hline
\textbf{Method}\qquad\qquad & \qquad $e$ \qquad\qquad &\qquad $m^2$ \\
\hline
DMRG~\cite{Huang2024}\qquad\qquad & \qquad -0.5503\qquad\qquad &\qquad 0.0433 \\
CG iPEPS~\cite{Chi2022}\qquad\qquad & \qquad -0.5520\qquad\qquad &\qquad 0.0541 \\
TL iPEPS~\cite{Naumann2026}\qquad\qquad & \qquad -0.5516\qquad\qquad &\qquad 0.0246 \\
VMC~\cite{Heidarian2009}\qquad\qquad & \qquad -0.5470\qquad\qquad &\qquad  0.0350 \\
\textbf{this work (extra. $1/D$)}\qquad\qquad & \qquad \textbf{-0.5517}\qquad\qquad &\qquad \textbf{0.0358} \\
\textbf{this work (extra. $1/\xi$)}\qquad\qquad & \qquad \textbf{-0.5511}\qquad\qquad &\qquad \textbf{0.0303} \\
\hline
\end{tabularx}
\end{table}

\begin{figure}[htbp]
\centering 
\includegraphics[width=0.95\columnwidth]{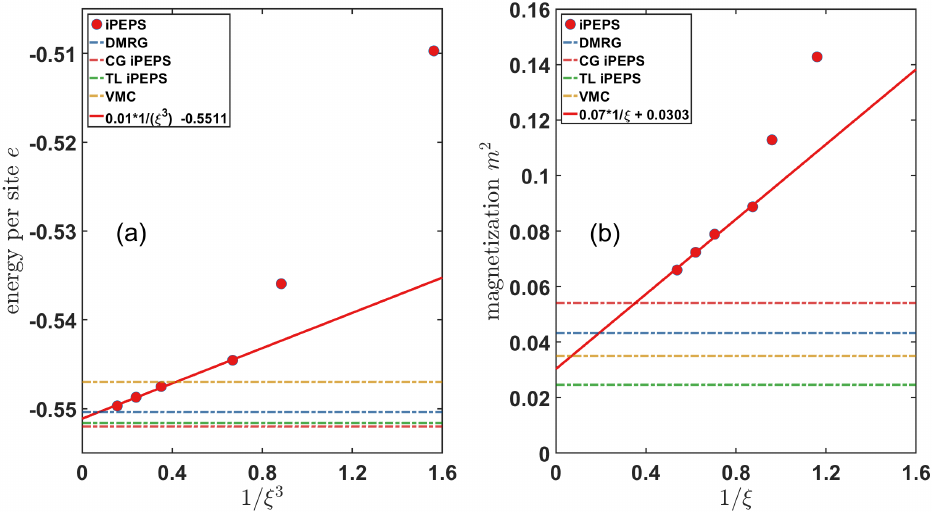}
\caption{Finite correlation length scaling for the triangular Heisenberg model. 
The ground state energy per site $e$ and magnetic order $m^2$ are fitted linearly versus $1/\xi^3$ and $1/\xi$, respectively, where $\xi$ is the maximal correlation length along $x$ direction. 
The data with $D=4,5,6$ and $7$ are used in the fits.
}
\label{Fig:finite_corr_scaling}
\end{figure}

The same data with finite correlation length scaling is shown in Fig.~\ref{Fig:finite_corr_scaling}. Following Ref.~\cite{Rader2018,Corboz2018}, we have fitted the energy density linearly in $1/\xi^3$, and magnetization $m^2$ linearly in $1/\xi$. (Notice that, since the effective square lattice does not have $C_4$ lattice rotation symmetry, the maximal correlation length in the $x$ and $y$ directions are not exactly the same, but provides very similar fits. For clarity, all correlation length/transfer matrix spectrum are along the $x$ direction unless otherwise explicitly stated.) As one can see from Fig.~\ref{Fig:finite_corr_scaling}, the quality of the fit depends on the data set being used, and therefore is not as good as the finite bond dimension scaling. In fact, for certain $J_2$ the maximal correlation length can even decrease with $D$. Similar behavior has also been observed in the highly frustrated regime of the $J_1$-$J_2$ model on the square lattice~\cite{Hasik2021}. One possible explanation is that the bond dimension is too small to enter the $\xi$-scaling regime. Thus, in the following, we shall rather use finite bond dimension scaling.

The consistency across a broad range of independent approaches validates our unrestricted iPEPS setup for handling the triangular lattice. Having established this benchmark, we next turn to the more challenging $J_2>0$ regime and to our search for a potential quantum spin liquid phase.

\subsection{The $120^\circ$ Néel state}

As mentioned in the introduction, it is now well accepted that the $120^\circ$ N\'eel states extends to a finite $J_2$, although results based on state-of-the-art PEPS methods have not been reported. Here our goal is to study how the magnetic order disappears with increasing $J_2$, therefore entering a non-magnetic phase.

Similar to the $J_2=0$ case, for each finite $J_2$ and bond dimension $D$, we use the same setup shown in Fig.~\ref{Fig:PEPS_setup} to variationally optimize the PEPS tensor. We find that the results are relatively easy to converge with environment bond dimension $\chi$. For a fixed bond dimension $D$, we observe that the energy density $e$ increases monotonically with increasing $J_2$, reflecting the enhanced frustration introduced by the next-nearest-neighbor interactions. While the magnetic moment decreases with $J_2$, it remains always nonzero at finite $D$.

\begin{figure}[htbp]
\centering 
\includegraphics[width=0.95\columnwidth]{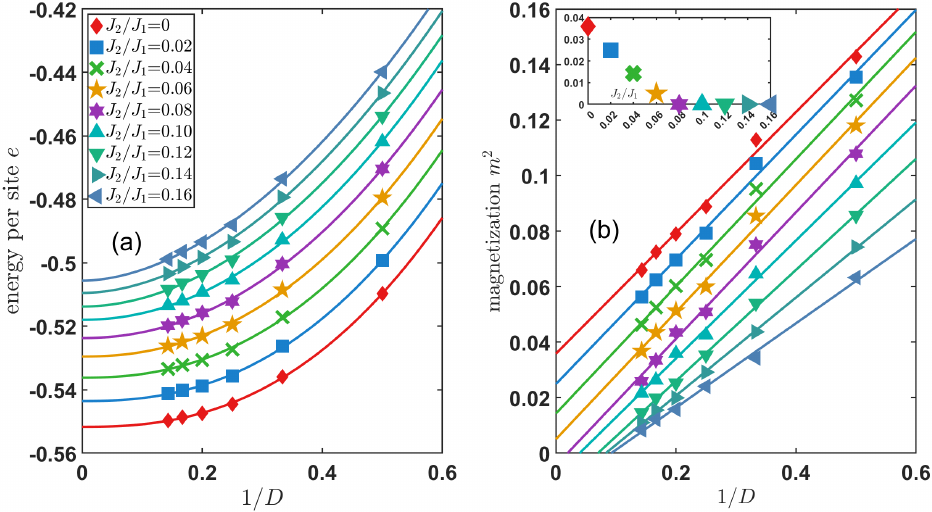}
\caption{
Finite bond dimension scaling for ground state observables of $J_1$-$J_2$ model. To extrapolate observables to $D\to\infty$, we have used the same power-law fit for energy density in (a), and linear fit for magnetization $m^2$ in (b). The extrapolated values for magnetization is shown in the inset of (b), showing a non-magnetic phase exists for $J_2\ge 0.08$.
} 
\label{Fig:finite_D_scaling_J2}
\end{figure}

We have performed a systematic finite‑$D$ extrapolation as shown in Fig.~\ref{Fig:finite_D_scaling_J2}.  For each $J_2$, we show the finite bond dimension scaling using the same scaling function as in Fig.~\ref{Fig:finite_D_scaling}. Note that we have considered here the scaling of $m^2$ instead of $m$, following Ref.~\cite{Hasik2021}. One can see that, with increasing $D$, the energy and magnetization for each $J_2\in [0, 0.16]$ are decreasing smoothly. We find that the extrapolated onsite magnetic moment vanishes around $J_2\approx 0.08$, shown in the inset of Fig.~\ref{Fig:finite_D_scaling_J2}. The disappearance of long‑range magnetic order signals a continuous transition from the $120^\circ$ N\'eel state to a nonmagnetic, highly entangled QSL phase.

Since we expect a power law behavior of the (connected) spin-spin correlation function in ordered magnetic phases, i.e. a diverging correlation length, it is tempting to use the finite correlation length scaling for nonzero $J_2$. However, we find that the maximal correlation length extracted from the transfer matrix (TM) remains small, typically $\xi<1.9$, even for the largest $D$. This suggests that $D$ is still not large enough to be in the finite-$\xi$ scaling regime. Note that relatively small finite-$D$ correlation lengths (typically $3< \xi<9$ for $D=7$) have also been observed on the square lattice~\cite{Hasik2021}.

While the zero magnetization indicates a spin liquid phase, the unrestricted PEPS simulation itself does not reveal much about the nature of the spin liquid. Before diving deeper into this question, we would like to complete the phase diagram study by investigating the second magnetic order phase shown in Fig.~\ref{Fig:phase_diagram}.

\subsection{Stripe AF state}

\begin{figure}[htbp]
\centering 
\includegraphics[width=0.95\columnwidth]{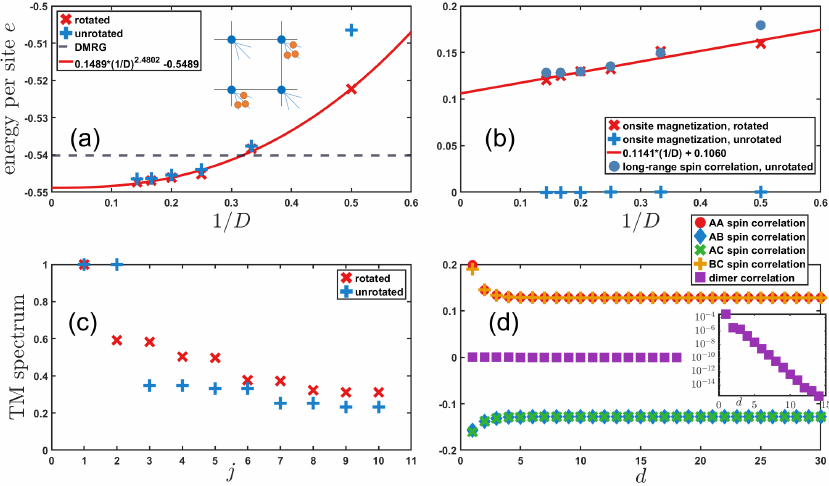}
\caption{
iPEPS investigation of the stripe phase at $J_2=0.5$. A comparison of unrotated PEPS and rotated PEPS results are shown in panels (a), (b), and (c), where red symbols indicate results obtained with the local unitary rotation, whereas blue symbols correspond to calculations without it.
Panels (a) and (b) show the finite‑$D$ scaling of the ground‑state energy and the staggered magnetization, respectively. Note the onsite magnetization of the unrotated PEPS vanishes. Panel (c) displays the transfer matrix spectrum of $D=7$ along the $x$ direction (results along the $y$ direction are similar). Panel (d) shows the correlation function of the unrotated PEPS at $D=7$, where spin correlations show long-range order and dimer correlations decay exponentially (inset). The magnetization order parameter of the unrotated PEPS is extracted from the long-range spin correlations, as shown in panel (b).
}
\label{Fig:stripe_phase}
\end{figure}

It is interesting to extend the unrestricted PEPS setup with one-site unit cell to larger $J_2$, and see how the magnetization further decreases at finite bond dimension $D$. From early studies it is known that the model will enter a stripe AF phase at sufficiently large $J_2$~\cite{Zhu2015,Hu2015}. {\it A priori}, it is not known how this phase could be revealed via such a PEPS simulation setup. On the other hand, although the stripe antiferromagnetic phase breaks one‑site translation symmetry on both the triangular and effective square lattices, an appropriate rotation of the local tensor restores an effective one‑site unit‑cell description for this phase~\cite{Hasik2021}, see inset of Fig.~\ref{Fig:stripe_phase}(a). (To be specific, it is $\pi$ rotation along $S^y$ on one of the sublattices on the mapped square lattice.) Motivated by the above, in this subsection we consider $J_2=0.5$, a value well within the stripe‑ordered regime, and perform iPEPS simulations under both rotated and unrotated PEPS ansatz with one site unit cell.

Figs.~\ref{Fig:stripe_phase}(a) and (b) show the finite‑$D$ scaling of the ground‑state energy and staggered magnetization, respectively, highlighting the contrast between the two approaches. A key observation is that, although the energies obtained from the rotated and unrotated PEPS are nearly degenerate~(Fig.~\ref{Fig:stripe_phase}(a)), their onsite magnetic moments differ dramatically. As shown in Fig.~\ref{Fig:stripe_phase}(b), the unrotated PEPS yields an almost vanishing magnetization (with magnitude around $10^{-3}$), whereas the rotated PEPS converges to a finite value characteristic of the stripe phase. Moreover, the orientation of the onsite magnetic moment extracted from the unrotated ansatz deviates significantly from the expected stripe pattern, providing further indications that, {\it apparently}, it fails to capture the correct magnetic order. On the contrary, the rotated PEPS ansatz correctly captures the pattern in this phase.

We note, however, that the energy and onsite magnetization are only local characterizations of PEPS. To gain deeper insight into the properties of the two ansatz, using well converged environment tensors from CTMRG, we have computed the leading part of the transfer matrix spectrum showing significant differences between the two. For the optimized unrotated PEPS, for $D=2,\cdots,7$, %with converged CTMRG environments, 
on both $x$ and $y$ directions the largest eigenvalue of the transfer matrix (see Fig.~\ref{Fig:stripe_phase}(c) for $D=7$ data) are two-fold degenerate, indicating that the state is a Schr{\"o}dinger cat state.
%with infinite correlation length.
This explains the onsite magnetization being close to zero in this ansatz. In contrast, for the rotated PEPS, the largest eigenvalue is non-degenerate and magnetic order sets in.
Note that the maximal correlation length extracted from the transfer matrix is still quite small (although it should diverge when $D\rightarrow\infty$).

One natural question related to the degeneracy of the leading eigenvalue is which type of correlation is long-ranged. Using CTMRG environment tensors, we have computed both spin-spin correlation and dimer-dimer correlation, defined as
\begin{align}
    & C^{ij}_{\mathrm{s}}(d)=\langle \mathbf{S}_{\mathbf{0},i}\cdot \mathbf{S}_{\mathbf{d},j}\rangle - \langle \mathbf{S}_{\mathbf{0},i} \rangle\cdot\langle \mathbf{S}_{\mathbf{d},j} \rangle,\\
    & C_{\mathrm{D}}(d)=\langle D^x_{\mathbf{0}} D^x_{\mathbf{d}} \rangle - \langle D^x_{\mathbf{0}} \rangle \langle D^x_{\mathbf{d}} \rangle,
\end{align}
where $i,j=A,B,C$ refer to the three sublattices, and without loss of generality we have restricted the dimer operator to $D^x_{\mathbf{r}}=\mathbf{S}_{\mathbf{r},A}\cdot\mathbf{S}_{\mathbf{r}+\mathbf{e}_x,A}$. The results of the correlation functions using the setups of Fig.~\ref{Fig:Z2corrlenfig}(a),(b) are shown in Fig.~\ref{Fig:stripe_phase}(d). While the dimer correlation show nicely exponential decay, the spin correlations are long-ranged. This behavior is also verified using periodic boundary condition (data not shown).

In fact, there are intricate structures in the spin correlation. We find that the long range correlation mainly comes from the $S^y$ component, while the $S^{x,z}$ components are vanishingly small. Interestingly, since the PEPS tensors are real, $\langle S^y\rangle$ is identically zero. Moreover, the sign of the long-range part in $C^{ij}$ depends on $i,j$. While the diagonal parts, i.e. $C^{AA}, C^{BB}, C^{CC}$, are positive, the off-diagonal parts $C^{AB}, C^{AC}$ are negative and $C^{BC}$ is positive. Note that all long-range spin correlations have close magnitudes. It is not hard to see that the structure in the correlations actually agrees with the stripe pattern. We have further compared the long-range correlation with the static magnetic moment of the rotated PEPS, finding a perfect agreement in Fig.~\ref{Fig:stripe_phase}(b).

Thus the breakdown of the injectivity in the optimized PEPS provides a clear signature of the stripe phase, ruling out a zero-magnetization spin liquid phase at $J_2=0.5$ (and extending to larger $J_2$). The pattern of long-range correlation further agrees with the magnetic moment orientation in the rotated PEPS, confirming the existence of the stripe phase. We would like to mention that long-range correlation is quite rare in variationally optimized PEPS. One may wonder why the CTMRG method still works in this case. We will discuss this point at the end of this work.

\section{Symmetric PEPS study of the QSL phase}
\label{Sec:QSL_phase}

One of the most intriguing physics in this model is the quantum spin liquid phase, for which various studies have led to different conclusions. It is interesting to note that a PEPS based method has not been applied to this phase yet. 
Here, we aim to fill this gap by investigating the properties of this phase using PEPS. Early works on the square~\cite{Poilblanc2017}, kagome~\cite{Niu2022}, and honeycomb~\cite{Lee2019} lattices have shown that an useful approach to reveal the nature of the spin liquid phase is to impose lattice and/or spin rotation symmetry in the PEPS ansatz. Here we follow the same route and systematically explore how different symmetry constraints influence the performance of the iPEPS ansatz, focusing on $J_2=1/8$.

\subsection{Implementing SU(2) symmetry on the coarse-grained lattice}

Since the spin liquid phase has vanishing magnetic order, one straightforward approach is to impose $\mathrm{SU}(2)$ symmetry into local tensors, so that the state is a $\mathrm{SU}(2)$ singlet by construction. To achieve that, using the same setup as in Fig.~\ref{Fig:PEPS_setup}, one need to specify the $\mathrm{SU}(2)$ representations carried by the virtual space $V$. Then we put the same onsite projector $P$ on each site which enforces the $\mathrm{SU}(2)$ fusion rules of four virtual spins into physical spins on each coarse-grained site, namely $P: V^{\otimes 4}\to (\frac{1}{2})^{\otimes 3}$ and, on each bond, the neighboring virtual spins are paired into a $\mathrm{SU}(2)$ singlet $|\Omega\rangle$. The PEPS tensor $\mathcal{A}$ in Eq.~\eqref{Eq:CG_PEPS} is finally obtained by absorbing two neighboring singlets $|\Omega\rangle$ into the onsite projector $P$. Since all onsite projectors and bond singlets are the same, this setup guarantees that the state is both $\mathrm{SU}(2)$ invariant and translationally invariant (on the square lattice to which the model is mapped).

Starting with a virtual space $V=\frac{1}{2}\oplus 0$, which corresponds to bond dimension $D=3$, it turns out that the optimized energy $e=-0.4561$ is far from that of the unrestricted PEPS simulations at the same bond dimension, $e=-0.4840$. Further enlarging the virtual space, we find that the energy of the $\mathrm{SU}(2)$ symmetric ansatz does not improve much (data not shown) and in fact the CTMRG procedure can occasionally have convergence issues. This points to the fact that this ansatz is too restrictive, breaking key spatial symmetries of the triangular lattice, including its sixfold rotational symmetry and mirror operations, thereby distorting the variational landscape and biasing the optimization away from the true ground state. Thus new approaches are needed.

\subsection{SU(2) symmetric PEPS on the original triangular lattice}

\begin{figure}
\centering
\includegraphics[width=0.95\columnwidth]{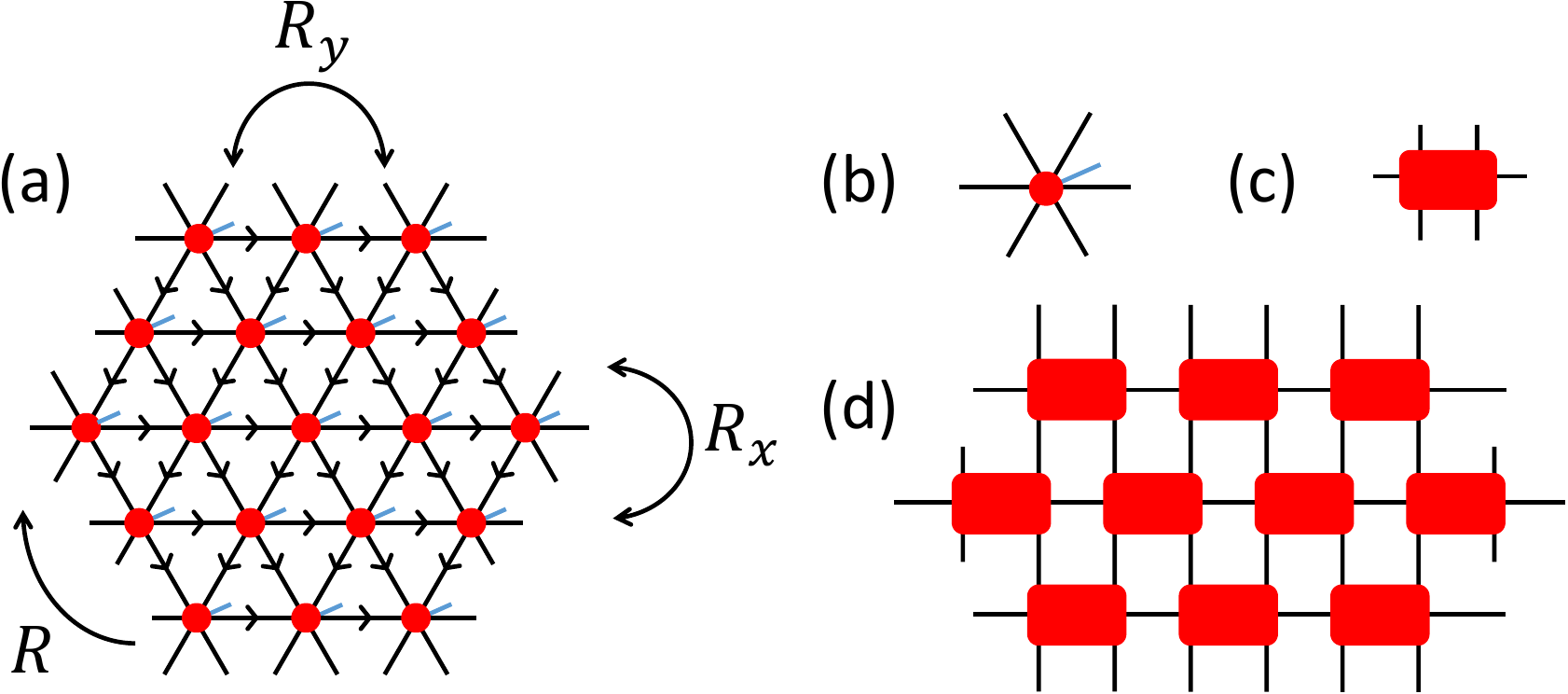}
\caption{
(a) $C_{6v}$ symmetric ansatz for the QSL phase, which naturally includes nearest-neighbor and long-range RVB states. The arrows indicate the singlet orientation. To contract the network for wave function norm, we note that the double tensor constrcted from the rank-7 PEPS tensor (shown in (b)) can be viewed as a ``dressed'' two-site gate (c). The norm network can be reshaped into matrix product operator composed of the two-site gates (d), and the contraction can be achieved by boundary MPS method.
}
\label{Fig:C6_ansatz}
\end{figure}

The first new approach we have considered is the symmetric PEPS on the original triangular lattice, where both $\mathrm{SU}(2)$ symmetry and the $C_{6v}$ point group symmetry can be realized by construction, see Fig.~\ref{Fig:C6_ansatz}. Similarly to the above construction of the SU(2) symmetric tensor on the effective square lattice, we specify the SU(2) representations on each virtual index, and then on each site we have a SU(2) projector $P$ from {\it six} virtual spins to the physical spin-$1/2$, and on each bond we require neighboring spin-$1/2$'s paired into a singlet $|\Omega\rangle$. To realize the lattice rotation symmetry and reflection symmetry, we have followed Ref.~\cite{Mambrini2016} to classify the projectors with the generators of $C_{6v}$ group, including $60^\circ$ rotation and reflection $R_x(R_y)$ along $x(y)$ direction. For the bond singlet, we have chosen the specific bond orientation of Fig.~\ref{Fig:C6_ansatz}(a) which respects the lattice point group symmetry. Note that, using the $\mathbb{Z}_2$ gauge symmetry of the local tensor, which originates from the center of the $\mathrm{SU}(2)$ group, one can flip the bond orientation so that lattice rotation symmetry becomes manifest.

% physical content for C6.
The symmetric PEPS defined on the triangular lattice has transparent and interesting physical content. With virtual space $V=1/2\oplus 0$, which corresponds to bond dimension $D=3$, one can represent the nearest-neighbor resonating valence bond (RVB) state exactly~\cite{Schuch2012}. Indeed, restricting the projector $P$ to the configuration with one virtual spin-1/2 and five virtual spin-0, the nearest-neighbor RVB state is uniquely realized in the $A_1$ representation of the $C_{6v}$ point group. Interestingly, the nearest-neighbor RVB state is the very first instance of quantum spin liquid state~\cite{Anderson1973}, and has been shown to be fully gapped (on non-bipartite lattices)~\cite{Poilblanc2012}. Further, by introducing configurations with either three or five virtual spin-$1/2$ one can realize, by quantum teleportation, long-range singlets in this $D=3$ ansatz~\cite{Chen2018}, therefore providing a simple way to realize a long-range RVB state.

% method for C6.
Although the physical picture is transparent, the computational method for PEPS on triangular lattice is less developed (however, see also Ref.~\cite{Naumann2026} for very recent methodology developments). Here to solve this issue, we observe that the double tensor in the wave function norm of PEPS can be viewed as a two-site gate defined on the virtual space, with two additional virtual indices on the left and right (see Fig.~\ref{Fig:C6_ansatz}(c)). With this view, the transfer operator in the norm network can be seen as composed by matrix product operator with two site gates (see Fig.~\ref{Fig:C6_ansatz}(d)), in contrast to the PEPS on the square lattice. Thus one can use the standard boundary matrix product state approach to find the top and bottom fixed-point of the transfer operator, achieving the contraction for the norm. The computation of physical observables and correlation length is then straightforward.

Due to the relatively high computational cost, especially in evaluating the expectation value of the $J_2$ term, 
we are restricted to a bond dimension of $D=3$ for this ansatz. Optimizing the variational parameters in the onsite projector, we have slightly improved the energy of the NN RVB $e=-0.4929$ to $e=-0.4959$ for the long-range RVB, which is only $1.2\%$ higher than the VMC estimate $e=-0.5020$ of the U(1) Dirac SL~\cite{Iqbal2016}. Note that, both values are obtained with environment bond dimension $\chi=36$, and we found enlarging $\chi$ turns out to be prohibitive due to the $J_2$ term.
We have further computed the correlation length and found a finite value (with maximal correlation length slightly increased from $2.63$ for NN RVB to $2.77$ at $\chi=36$), which suggests that the optimized $D=3$ long-range RVB phase is still gapped.

\subsection{$\mathbb{Z}_2$ symmetric ansatz on the effective square lattice}
\label{Subsec:Z2}

The $D=3$ rank-7 tensor suggests a gapped $\mathbb{Z}_2$ spin liquid, but is limited to small bond dimension. Guided by this result, and to systematically improve the energetics, we go back to the coarse-grained lattice, while restricting the essential symmetry to the $\mathbb{Z}_2$ gauge symmetry. This is a reasonable choice since some symmetry breaking orders, including magnetic order and lattice symmetry breaking have been shown to disappear under finite bond dimension scaling (see Sec.~\ref{Sec:unrestricted_PEPS}) and are therefore artifacts of finite-$D$ PEPS. Note that, with $\mathbb{Z}_2$ gauge symmetry implemented, the setup is biased towards $\mathbb{Z}_2$ topological order. However, the $\mathbb{Z}_2$ topological order is not guaranteed by construction. One would still need to check if all correlation lengths, including those in the flux sectors, remain finite to make sure the topological excitations (anyons) are deconfined~\cite{Duivenvoorden2017,Iqbal2018,Iqbal2021}. Also, one would further need to evaluate the energetics to make a fair comparison. On the other hand, if certain correlation length diverges or anyon condensation appears at finite bond dimension, such results would strongly rule out $\mathbb{Z}_2$ topological order.

\begin{figure}[htbp]
\centering 
\includegraphics[width=0.95\columnwidth]{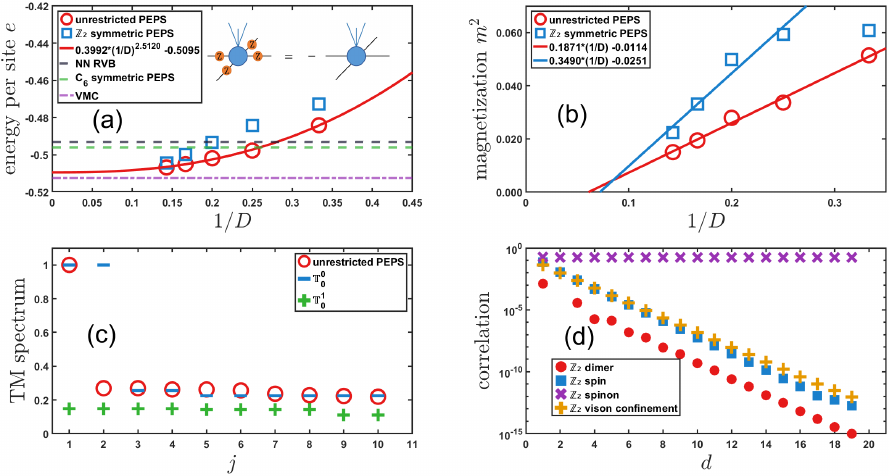}
\caption{(a) Comparison of the ground-state energies - plotted versus $1/D$ - of the $\mathbb{Z}_2$-symmetric PEPS and the unrestricted PEPS on the effective square lattice. The energies of the ($C_{6v}$-symmetric) $D=3$ NN and long-range RVB states, and the VMC energy are also shown.
(b) Onsite magnetization for unrestricted PEPS and $\mathbb{Z}_2$ symmetric PEPS, both extrapolated to a vanishing value at large $D$ limit. For the $\mathbb{Z}_2$ case, $D=3$ is excluded in the linear fit. (c) Transfer matrix spectrum of unrestricted and $\mathbb{Z}_2$ symmetric PEPS with $D=7$. For the former, the leading eigenvalue is unique. For the $\mathbb{Z}_2$ case, the leading eigenvalue in the flux free sector is two fold degenerate in magnitude. The spectrum of $\mathbb{T}^1_0$ is normalized with the largest eigenvalue of $\mathbb{T}^0_0$, and only the magnitude is plotted. (d) Correlation functions of $\mathbb{Z}_2$ symmetric PEPS. While spin and dimer correlations decay exponentially, long-range order in spinon correlation is observed. In contrast, the norm of the wavefunction with a vison line inserted shows an exponential decay with the length of the vison line, suggesting vison confinement.
} 
\label{Fig:Z2_ansatz}
\end{figure}

Since the $\mathbb{Z}_2$ gauge symmetry originates from the center of the physical $\mathrm{SU}(2)$ symmetry, to implement the $\mathbb{Z}_2$ gauge symmetry, we require the local tensor $\mathcal{A}$ in Eq.~\eqref{Eq:CG_PEPS} satisfies the following equation:
\begin{equation}
    \left(Z\otimes Z\otimes Z\otimes Z\right) \circ \mathcal{A} = -\mathcal{A},
\label{Eq:Z2_gauge}
\end{equation}
which means acting on all virtual indices simultaneously with gauge symmetry generator $Z$ leaves the tensor invariant up to a sign, see inset of Fig.~\ref{Fig:Z2_ansatz}(a). Here $Z$ is a reducible representation of the $\mathbb{Z}_2$ generator, which we have chosen to be a diagonal matrix with nonzero elements $\pm 1$. The minus sign in Eq.~\eqref{Eq:Z2_gauge} comes from the fact that the center of $\mathrm{SU}(2)$ is generated by $\mathrm{exp}(i2\pi S^z)$, whose action on the three physical spin-$1/2$s will result in a minus sign.

It still remains to determine the $Z$ matrix explicitly. Again since the $\mathbb{Z}_2$ comes from $\mathrm{SU}(2)$, we have started with bond dimension $D=3$, and for each bond dimension, we consider all the possible $\mathrm{SU}(2)$ multiplet structure that is compatible, and then take the center to get the $Z$ matrix. For instance, with $D=3$, the only virtual space is $V=1/2\oplus 0$, for which $Z=\mathrm{diag}(-1,-1,1)$. For $D=5$, two possible $\mathrm{SU}(2)$ multiplets are $V=1/2\oplus 1/2\oplus 0$ and $V=1/2\oplus 1$, and thus $Z=\mathrm{diag}(-1,-1,-1,-1,1)$ and $Z=\mathrm{diag}(-1,-1,1,1,1)$, respectively. With $Z$ determined, one can then use Eq.~\eqref{Eq:Z2_gauge} to compute the basis for this virtual space, and the local tensor $\mathcal{A}$ is then taken to be a linear superposition of the basis.

With this approach, we have enumerated all the tensor classes up to bond dimension $D=7$. For a given bond dimension, we carry out independent optimization for the coefficients in each class, and take the lowest energy as the best result. In this way, we can systematically assess which symmetry structure provides the most accurate and energetically favorable description of the QSL ground state.

The results with $\mathbb{Z}_2$ symmetric PEPS ansatz are shown in Fig.~\ref{Fig:Z2_ansatz}. The optimal virtual space we have identified is characterized by gauge symmetry generator 
\begin{align}
&Z_{D=3}=\mathrm{diag}(-1,-1,1),\nonumber  \\
&Z_{D=4}=\mathrm{diag}(-1,-1,1,1), \nonumber\\
&Z_{D=5}=\mathrm{diag}(-1,-1,-1,-1,1), \\
&Z_{D=6}=\mathrm{diag}(-1,-1,-1,-1,1,1), \nonumber\\
&Z_{D=7}=\mathrm{diag}(-1,-1,-1,-1,1,1,1),\nonumber
\end{align} 
(and all permutations of the characters) for $D=3,4,5,6,7$, respectively. Comparing the variational energies (see Fig.~\ref{Fig:Z2_ansatz}(a)), we find that the $\mathbb{Z}_2$ symmetric ansatz has an energy comparable to that of the unrestricted PEPS ansatz, both of which are lower than the VMC estimate $e=-0.5020$ of the (plain) U(1) Dirac spin liquid (improved to $-0.5124$ after applying two Lanczos steps)~\cite{Iqbal2016}. Moreover, the energy difference between the two PEPS ansatz decreases with bond dimension $D$, suggesting that the $\mathbb{Z}_2$ ansatz captures the true ground state. 
Since only $\mathbb{Z}_2$ gauge symmetry is imposed, a finite magnetization cannot be avoided, whose value we have compared with the unrestricted ansatz in Fig.~\ref{Fig:Z2_ansatz}(b). Under finite bond dimension scaling, both magnetizations tend to vanish in the $D\to\infty$ limit. It is interesting to note that, at finite $D$, the magnetic moment in both ansatz shows a $120^\circ$ order.

We now further analyze the properties of the $\mathbb{Z}_2$ symmetric ansatz to identify the nature of the variational optimum. To confirm or deny $\mathbb{Z}_2$ topological order, we compute the transfer matrix spectrum, which can be conveniently obtained from the CTMRG environment tensors (see Fig.~\ref{Fig:Z2_ansatz}(c)). Due to the $\mathbb{Z}_2$ gauge symmetry, we can further insert $\mathbb{Z}_2$ flux line in either bra or ket layer of the wave function norm network, to construct the flux sectors~\cite{Schuch2013}. We denote the transfer matrix in the four flux sectors as $\mathbb{T}^{\alpha}_{\beta}$, where $\alpha,\beta=0 (1)$ represents zero ($\mathbb{Z}_2$) flux line inserted in the bra and ket layer, respectively.

It turns out that, for all bond dimension $D$, the largest eigenvalue of the transfer matrix in zero-flux sector $\mathbb{T}^{0}_{0}$ is two-fold degenerate in absolute value, with opposite signs. This signifies that the boundary MPS constructed with CTMRG environment tensors is not injective, and certain long-range order exists in the optimized PEPS. To find out which observable is long-range ordered, we have computed explicitly a few cases, including spin, dimer, and spinon correlation functions, using the CTMRG environment tensors, as shown in Fig.~\ref{Fig:Z2corrlenfig}. Note that, here the spinon is a topologically nontrivial quasiparticle, the construction of which requires an operator $X$ acting on the virtual indices. The operator $X$ further needs to satisfy the relation $XZ=-ZX$, and is not unique~\cite{Duivenvoorden2017}. 
For instance, one reasonable form of $X$ for $V=1/2\oplus 0$ is given by 
$X=\begin{pmatrix}
0 & 0 & 1\\
0 & 0 & 0\\
1 & 1 &0
\end{pmatrix}$, which anticommutes with $Z_{D=3}$.

We found that, while the spin and dimer correlations show exponential decay, the spinon correlator (inserting two $X$ operators on virtual bonds at increasing distance, see Fig.~\ref{Fig:Z2corrlenfig}(c)) shows perfect long-range correlation, as seen in Fig.~\ref{Fig:Z2_ansatz}(d). This indicates the spinon is condensed~\cite{Duivenvoorden2017} in the variational optimum. Interestingly, the spinon correlation alternates in sign with distance, nicely matching the sign of the leading eigenvalues.

To get the behavior of vison excitations, we explore the spectrum with flux line inserted. We find that while the leading eigenvalues of $\mathbb{T}^1_1$ are the same as those of $\mathbb{T}^0_0$ with two fold degeneracy in magnitude, the leading eigenvalue of $\mathbb{T}^1_0$ and $\mathbb{T}^0_1$ are strictly smaller than that of $\mathbb{T}^0_0$. Further, $\mathbb{T}^1_0$ and $\mathbb{T}^0_1$ show the same spectrum, with leading eigenvalues showing multi-fold degeneracy in magnitude (see Fig.~\ref{Fig:Z2_ansatz}(c)). This behavior is consistent with breaking $\mathbb{Z}_2\times \mathbb{Z}_2$ symmetry of the transfer operator to $\mathbb{Z}_2$ symmetry in the boundary state, where simultaneously acting with $Z$ on the bra and ket layer leaves the boundary state invariant. However, computing the norm of the wave function with flux line inserted (see Fig~\ref{Fig:Z2corrlenfig}(d)), we found the norm is decreasing exponentially with the length of the flux line which suggests that the remaining $\mathbb{Z}_2$ symmetry in the boundary is in fact {\it spontaneously} broken. This is different from the approach in Ref.~\cite{Iqbal2018}, where the boundary state is initialized with a random tensor thus breaking {\it explicitly} the symmetry (and leading to different $\mathbb{T}^1_1$ and $\mathbb{T}^0_0$ spectra). This suggests the visons are in fact confined~\cite{Duivenvoorden2017} in contrast to e.g. the Toric Code with potentially small deformation for which the state with an inserted vison flux (obtained e.g. from a vison pair brought to infinite separation) is well defined (i.e. normalizable which we have checked explicitly, data not shown) and corresponds to the $\mathbb{Z}_2$ topological partner~\cite{Iqbal2018}. Our results are consistent with the conventional scheme that the condensation of a given anyon type results in the confinement of the dual particles (and of all other anyons - if any -  which have nontrivial mutual statistics with the condensed type)~\cite{Bais2009}.

Thus the $\mathbb{Z}_2$ topological order is ruled out in this case. From early theoretical studies, it is known that one possibility of bosonic spinon condensation is to induce magnetic order, which on the triangular lattice would be a $120^\circ$ order~\cite{Sachdev1992} and is consistent with the order we observed at finite bond dimension for the $\mathbb{Z}_2$ symmetric PEPS. Interestingly, we have checked explicitly in the $D=3$ case, that degeneracy in the flux free sector is not an isolated point in the variational manifold. In fact, the degeneracy does not appear at the initial stage of the variational optimization but, rather, appears systematically after a finite number of optimization steps, showing that the degeneracy is not an artifact of our symmetric ansatz. Thus we conclude that the $\mathbb{Z}_2$ topological order vanishes at the variational optimum.

The above results for symmetric PEPS and the comparison to results from unrestricted PEPS, provide hints on the nature of the quantum spin liquid phase in this model. The fact that long-range correlation in the optimized $\mathbb{Z}_2$ symmetric PEPS is associated to spinons, instead of ordinary local observables, is compatible with the unrestricted PEPS simulation, for which we always find that the leading eigenvalue of the transfer matrix is non-degenerate at finite $D$ and $J_2=1/8$. Indeed, spinon excitations can be constructed in the former, while not in the latter. Despite the fact that the $\mathbb{Z}_2$ ansatz is biased towards $\mathbb{Z}_2$ topological order, we instead find spinon condensation, which indicates absence of $\mathbb{Z}_2$ topological order. Note also that, although the $C_{6v}$ ($D=3$) symmetric ansatz carries $\mathbb{Z}_2$ topological order, its variational energy is still too high to be truly competitive and to legitimate such a state as a valid QSL candidate. Hence, the above results suggest $\mathbb{Z}_2$ QSL is unlikely to be realized in this model.

\section{Conclusion}
\label{Sec:conclusion}

This work systematically investigates the phase diagram of the frustrated $J_1$-$J_2$ antiferromagnetic Heisenberg model on the triangular lattice using an advanced projected entangled pair states approach. Our central finding is the revelation of a rich phase diagram comprising a $120^\circ$ N\'eel order, a quantum spin liquid phase, and a stripe phase.

First, the successful benchmarking of the $J_2 = 0$ case (Fig.~\ref{Fig:finite_D_scaling}, Table~\ref{Tab:benchmark}) provides a solid proof of the accuracy of our method. Subsequent scanning of the ground state energy and magnetic order parameter versus $J_2/J_1$ (Fig.~\ref{Fig:finite_D_scaling_J2}) clearly delineates the phase boundaries. In particular, the abrupt collapse of magnetic order around $J_2/J_1 \approx 0.08$ provides compelling evidence for the existence of a spin liquid phase in a finite intermediate region. This finding is consistent with prior variational Monte Carlo~\cite{Iqbal2016} and DMRG studies~\cite{Zhu2015,Hu2015}.

More importantly, our work delves into the crucial role of symmetry in the tensor network characterization of complex quantum phases. E.g. the investigation of the stripe phase at $J_2/J_1=0.5$ (Fig.~\ref{Fig:stripe_phase}) revealed interesting features. Calculations performed on the effective (coarse-grained) square lattice were carried out in two ways, either by implementing the correct real-space modulation of the stripe phase (i.e. by doubling the unit cell) or by assuming translation invariance. In the first case, the correct magnetic pattern together with a good variational energy were obtained. In contrast, the second realization was giving an (almost) degenerate Schr\"odinger cat state, with vanishing magnetization, from which the correct magnetic order parameter could only be extracted from the analysis of the long-range spin-spin correlations. This highlights the importance of characterizing variationally optimized PEPS through non-local observables (e.g. correlation functions), besides conventional local ones.

Furthermore, our study of the QSL at $J_2/J_1=0.125$ (see e.g. Fig.~\ref{Fig:C6_ansatz} and Fig.~\ref{Fig:Z2_ansatz}) revealed another profound insight. An ansatz with only $\mathbb{Z}_2$ symmetry turned out to provide a QSL candidate with a remarkably good variational energy (in the limit of bond dimension $D \rightarrow \infty$). This fact seems to suggest that the spin liquid phase exhibits $\mathbb{Z}_2$ topological order (and is gapped). However, a careful examination of the properties of the $\mathbb{Z}_2$ PEPS revealed in fact that spinon excitations are condensed and vison excitations are confined, hence precluding the occurrence of topological order. Interestingly, these results do not exclude the proposed $\mathrm{U}(1)$ Dirac spin liquid phase~\cite{Iqbal2016} although no direct evidence of this phase is given. We believe this type of approach may also be useful for other highly debated models in frustrated magnetism, e.g., the $J_1$-$J_2$ model on square lattice and kagome lattices. As $\mathbb{Z}_2$ gauge symmetry is much easier to encode into local tensors than constructing a genuine (2+1)D critical spin liquid, our approach provides a way to prove or rule out the existence of gapped spin liquid in these models.

Technically, one interesting feature of our work is the finding of stable numerical PEPS calculations even when the leading eigenvalues of the transfer matrix are degenerate. Indeed, both the spin long-range order in the stripe phase and the spinon long-range order in the $\mathbb{Z}_2$ symmetric ansatz for QSL correspond to such a case. To our knowledge, this behavior is quite uncommon in numerical optimization with PEPS, except the cases where the PEPS is constructed analytically~\cite{Rispler2015,Duivenvoorden2017,Iqbal2018} or the target phase diagram is known beforehand~\cite{Iqbal2021}. One may wonder why CTMRG still works quite well for this case, despite that the fixed point is degenerate. We notice that the CTMRG typically starts with an initial environment tensor constructed from the PEPS site tensor itself, hence inheriting the internal symmetries. Then, it is reasonable that the fixed point of the corner transfer operator also carries this symmetry, which leads to a Schr\"odinger cat like environment. The degeneracy in the transfer matrix spectrum is then enforced by symmetry. For the stripe phase, to remove the degeneracy and reduce entanglement, we need to use environment tensors obtained with an enlarged unit cell, insuring the correct type of magnetic ordering. For the $\mathbb{Z}_2$ case, we need to break the virtual symmetry in the initial environment (which has no effect for a truly topological phase). Neither of these procedures are natural from a numerical optimization perspective, given that the looked-for quantum phase is {\it a priori} unknown.

Revealing the exact nature of the QSL phase using the PEPS methodology still remains. In recent work, excitation spectra and dynamical structure factors for this model have been computed using various methods, including ED~\cite{Wietek2024}, VMC~\cite{Budaraju2025}, and MPS based methods~\cite{Hu2019,Sherman2023,Drescher2023,Drescher2026}. In particular, monopole excitations have been explicitly studied using VMC~\cite{Budaraju2025}, providing evidence for the $\mathrm{U}(1)$ Dirac spin liquid. Conducting the same exploration with PEPS should be very fruitful.

\section*{Acknowledgement}
We thank Shenghan Jiang, Yi Xu for useful discussions.
This work is supported by National Natural Science Foundation of China (Grants No.~12304186, No.~12447107), Guangdong Basic and Applied Basic Research Foundation (Grant No.~2024A1515013065), and Quantum Science and Technology - National Science and Technology Major Project (Grant No.~2021ZD0302100). 
W.-L.T. is supported by the Center of Innovation for Sustainable Quantum AI (JST Grant Number JPMJPF2221) and JSPS KAKENHI Grant Number JP25H01545 and JP26K17054.

% \section*{Data availability}
% Data, data analysis, and simulation codes are available upon reasonable request on Zenodo.

\bibliography{draft}

\end{document}